\documentclass[twocolumn,showpacs,prl]{revtex4}

\usepackage{epsfig}
\usepackage{amsmath}
\usepackage{graphicx}
\usepackage{dcolumn}
\usepackage{amsmath}
\usepackage{latexsym}

\begin{document}

\title{Itinerant Ferromagnetism in the electronic localization limit}
\author{N. Kurzweil}
\author{E. Kogan}
\author{A. Frydman}
\address{The Department of Physics, Bar Ilan University, Ramat Gan 52900, Israel}

\begin{abstract}
We present Hall effect, $R_{xy}(H)$, and magnetoresistance,
$R_{xx}(H)$, measurements of ultrathin films of Ni, Co and Fe with
thicknesses varying between 0.2-8 nm and resistances between
1 M$\Omega$ - 100 $\Omega.$ Both measurements show that films having resistance above a critical value, $R_{C}$, (thickness below a critical value, $d_{C}$) show no signs for ferromagnetism. Ferromagnetism appears only for films with $R<R_{C}$, where $R_{C}$ is material dependent.  We raise the possibility that the reason for the absence of spontaneous magnetization is suppression of itinerant ferromagnetism by electronic disorder in the strong localization regime.
\end{abstract}


\date{\today}

\maketitle

Ferromagnetism in the transition metals (Fe, Co and Ni) relies, at
least partially, on the itinerancy of the conduction band
electrons. The balance between kinetic energy loss and
exchange energy gain leads to polarization of the band electrons and to
magnetic order. It is interesting to ask what happens when the
mobility of the electrons is strongly suppressed by disorder. In
the extreme case electrons can be localized with localization
length $\xi$, that may be of atomic scale. Under these conditions one may
expect itinerant ferromagnetism to be entirely suppressed since the
electronic functions do not overlap and exchange energy is no
longer relevant. This may be analogous to the situation in uniform disordered superconductors, where the superconductivity is suppressed by strong disorder as a result of disorder induced pair breaking (see for example  \cite{dynes, goldman, sasha}).

In order to achieve high enough disorder for significant localization in homogeneous metallic materials it is necessary to use very thin amorphous films having thicknesses of a few mono-atomic layers in which the sheet conductance can be of the order of $\frac{e^{2}}{h}$ or less. For this purpose the samples studied in this work were Ni, Co and Fe thin
films fabricated using "quench condensation" (evaporation on a cryo-cooled substrate).
 This technique allows to deposit sequential layers of ultrathin films and measure transport without thermally cycling the sample or exposing it to atmosphere. If a thin underlayer of Ge or Sb is pre-deposited before quench condensing a metal film, the underlayer wets the substrate thus enabling the growth of continuous ultrathin amorphous layers even at monolayer thickness \cite{strongin}. The underlayer being an insulator at low temperatures is assumed to have negligible effect on the electric properties of the metal \cite {strongin, goldman}.  The ability to study a single sample while driving it from strong localization to weak localization has been vastly used in the context of the superconductor-insulator transition. In the current work we used this technique to drive a ferromagnetic amorphous film from strong to weak localization. We studied  Ni, Co and Fe films having resistance in the range of 1 $M\Omega$ - 100 $\Omega$ and thicknesses of 0.2-8 nm. The films were evaporated on a 2nm thick layer of Sb or Ge, in a Hall bar geometry allowing 4 probe resistance and Hall effect measurements. Both sample resistance and thickness were monitored during the growth and the process was stopped at different evaporation stages. Room temperature AFM measurements of these films (performed in ambient after heating them up) reveal thickness roughness smaller than 1nm.

\begin{figure}
{\epsfxsize=3 in \epsffile{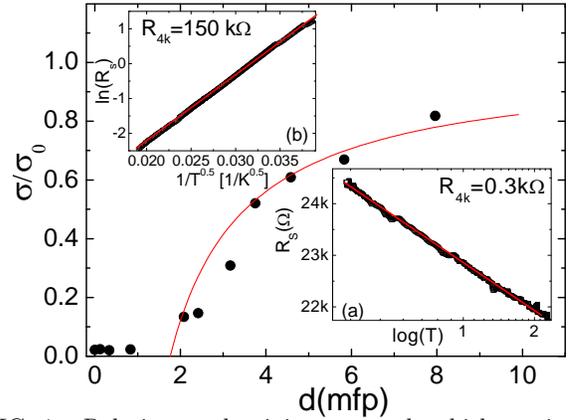}}
\vspace{-0.3cm}
\caption{ \small Relative
conductivity versus the thickness in mean free path units for a Fe
sample . The solid line is a fit to eq. 1. The insets show
resistance as a function of temperature in the weak (a) and strong
(b) localization regime (for 3.5 and 0.3 nm thick samples
respectively).} \label{particle}
\vspace{-0.3cm}
\end{figure}

Figure 1 shows the normalized conductance versus thickness of a Fe sample. The data is fitted the following expression extracted from the scaling theory of localization \cite{WL,future}:

\begin{equation}
    \sigma/\sigma_{0}=1-3A/(2k_{f}^{2}dl)ln[\tau_{\varphi}/3\tau],
\end{equation}

where $\sigma_{0}$ is the Drude conductivity, $k_{f}$ is Fermi wave length of the bulk, d is the thickness of the film and $\tau$ and $\tau_{\varphi}$ are the transport and dephasing time respectively. Eq. 1. is well known in the theory of weak localization. We have introduced a phenomenological factor A$\approx$3 which enables us to extend the fit well beyond the expected validity range into the strong disorder regime. The discussion on justification of this procedure will be presented elsewhere \cite{future}. The fitted relevant lengthscales: l-the mean free path and $l_{\varphi}$ - the dephasing length are listed in table 1.
\begin{table} \label{t1}
{\protect
\begin{ruledtabular}
\begin{tabular}{|c|c|c|c|c|c|}
\hline
$      $&$l$ (nm)&$l_{\varphi}$ (nm)&$R_{C}$ ($\Omega$)&  $d_{C}$ (nm)& $\mu$ $(\mu_{B})$\\ \hline
$\hspace{.15cm}Fe\hspace{.15cm}$    &  $0.4$ &       $46$       &     $80k$       &     0.5      &      2.2         \\ \hline
$\hspace{.15cm}Co\hspace{.15cm}$    & $0.29$ &       $21$       &     $10k$        &     0.8      &      1.7         \\ \hline
$\hspace{.15cm}Ni\hspace{.15cm}$    & $0.27$ &       $25$       &      $2k$        &     1.8      &      0.6         \\ \hline
\end{tabular}
\end{ruledtabular}
\vspace{-0.1cm}
} \caption{\small Mean free path, l, dephasing length,
$l_{\varphi}$, critical resistance and thickness for the appearance
of ferromagnetic signatures, $R_{C}$ and $d_{C}$,  and bulk magnetic
moments, $\mu_{B}$, for the Ni,Co and Fe films use in this study.}

\end{table}
The insets depict the resistance versus temperature curves for 0.3 nm and 3.5nm thick Fe film showing Efros-Shklovskii like hopping behavior and logarithmic dependence typical of weak localization behavior respectively. Extracting the localization lengths from such R(T) curves allows us to determine the crossover from strong localization $(\xi < L_{\varphi})$ to weak localization  $(\xi > L_{\varphi})$ at a sheet resistance of $\sim10$ $k\Omega$ and thickness of $\sim0.8$ nm.

In order to probe the magnetic state of the films we measured both Hall effect (HE) and magnetoresistance (MR) for each evaporation stage by applying a magnetic field perpendicular to the films. All presented results were obtained at T=4K. Hall resistivity in magnetic materials is combined of the ordinary part, $\rho_{0}$, observed in all normal metals due to the Lorentz force which is proportional to the magnetic field, H, and the extraordinary Hall effect (EHE), $\rho_{xy}$, which is proportional to the magnetization of ferromagnetic films, M, so that $\rho_{xy}\propto R_{EHE}\mu_{0}M$, $R_{EHE}$ being the extraordinary Hall effect coefficient. The two processes, which are believed to have the major contribution to the EHE, skew scattering and side jump, are both scattering processes, and the EHE is expected to depend on the longitudinal resistivity (which is also due to scattering) in the following way:
\begin{equation}
 R_{EHE}=a\rho_{xx}+b\rho_{xx}^{2}
 \end{equation}
 where $a$ is the skew scattering coefficient and $b$ is the side jump coefficient. Gerber et-al \cite{gerber} showed that for very thin films the dominant factor is skew scattering and $\rho_{xx}$ was found to be proportional to $\rho_{xy}$ for Ni films between 4 to 20 nm thick. For high magnetic fields the magnetization saturates and only the ordinary part contributes to the magnetic field dependence of the HE. Studying the EHE is an elegant way to measure magnetic properties of thin films using transport measurement, in particular, the saturation magnetization can be obtained by extrapolating the Hall resistance to zero field.

\begin{figure}
\vspace{-0.8cm}
{\epsfxsize=2.8 in \epsffile{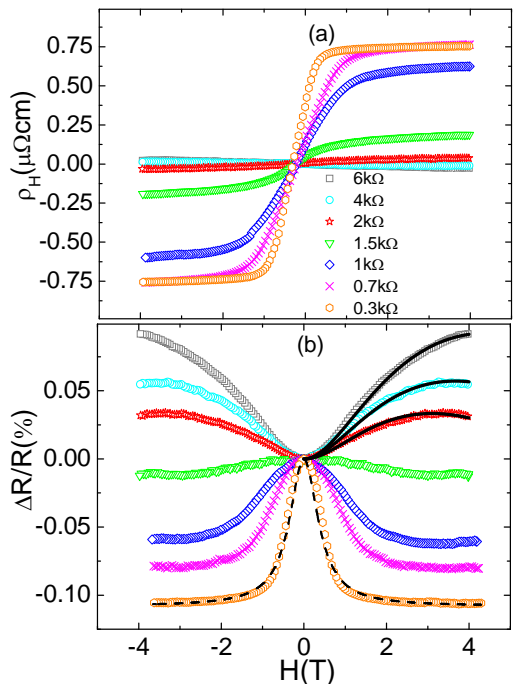}}
 \vspace{-0.9cm}
\caption{ Hall effect (a)
and magnetoresistance (b) measurements of sequential quench
condensed Ni films. T=4.2K. The dash line in the negative MR of the 0.3
$k\Omega$ stage is the fit for to the AMR phenomenological expression in
eq. 3. The solid lines in the positive MR curves show fits to the
anti weak-localization theory of ref \cite{chakravarty}  with $l$=0.265 nm,
$l_{\varphi}=25$ nm and $l_{so}$ equal to 20, 21.6 and 22.7 nm for
films with sheet resistance 6, 4, and 2 $k\Omega$ respectively.
\small} \label{particle}
\vspace{-0.3cm}

\end{figure}

Fig 2a. shows the Hall effect of sequential quench condensed layers
of Ni. Down to a sheet resistance of 2 k$\Omega$ (1.8 nm thick) only
the ordinary Hall effect is observed. As more material is added and
the resistance further decreased an EHE contribution develops and
increases with increasing thickness. This defines a critical
resistance, $R_{C}$, and critical thickness, $d_{C}$, for the
appearance of the EHE. Fig 3 shows both the longitudinal
resistivity, $\rho_{xx}$, and the transverse (Hall) resistivity
(determined from extrapolating to H=0), $\rho_{xy}$, as a function of thickness for a typical Ni
sample. $\rho_{xx}$ decreases monotonously while $\rho_{xy}$ is zero
for the thinnest films. For films thicker than 2 nm $\rho_{xy}$
increases with thickness until at d=4-5 nm it changes its trend. For
thick enough samples (d$>$4 nm) $\rho_{xy}\propto\rho_{xx}$ in
accordance with \cite{gerber}. If the $R_{EHE}\propto\rho_{xx}$
expression is extrapolated to smaller d, our results indicate that
the macroscopic magnetization, M, is zero for very resistive samples
in the strong localization regime, it grows in an intermediate
regime until it reaches saturation for films having sheet resistance
below 500 $\Omega$.

Interestingly, the magnetoresistance also undergoes a unique crossover at the same critical region. For high resistance (small thickness) a positive magnetoresistance is observed, whereas for $R<R_{C}$ ($d>d_{C}$) the MR changes sign and becomes negative, saturating at fields above a saturation field, $H_{S}$. This is seen in figure 2b which shows that the MR exhibits a change of MR sign for resistances below 2 k$\Omega$ ($R<R_{C}$).

Similar behavior is observed in Co and Fe as well, however $R_{C}$ and $d_{C}$ vary from material to material. Table 1 summarizes the critical thicknesses and resistances for which EHE appears and the MR changes sign for the three transition metals.

Just like the emergence of EHE, we interpret the MR sign change as a
signature for the appearance of ferromagnetism in the film. Positive
MR such as seen in our thin layers is also observed in quench
condensed non-magnetic layers such as Ag, where the MR remains
positive  for films with resistance as low as 100 $\Omega$. This
behavior is understood as being due to weak localization in the
presence of strong spin orbit. Since the films are ultrathin, one
can indeed expect strong Rashba spin orbit scattering \cite{rashba}
on the surface leading to weak anti-localization and positive MR.
The solid lines in fig 2b shows fits of our curves to the
spin-orbit weak localization expression \cite{chakravarty}. The spin
orbit length, $l_{SO} $, grows as the film thickens  as can be
expected if Rashba spin orbit is the dominant factor.

The crossover to negative MR occurs only in ferromagnetic materials. Dugaev \cite{dugaev} showed that weak anti-localization due to spin orbit interaction does not occur in ferromagnetic systems so that if ferromagnetism is present in a film, the MR should be always negative. Hence, the MR sign change is also an indication for the appearance of magnetization in the film. We have tried to fit the negative MR curve to the Dugaev expression \cite{dugaev} but could obtain reasonable fits only for very thin ($d<0.5 nm$) Fe layers \cite{future}. For thicker samples we find that the curves are well described by anisotropic magnetoresistance (AMR) typical to ferromagnetic films.  Since the field is applied perpendicular to the current direction one can expect a negative contribution that should depend quadratically on the angle between the current and the magnetization \cite{AMR}. We have fitted our curves to the following phenomenological expression that assumes quadratic dependance of $\Delta R$ on H for low fields and saturation for fields larger than the saturation field, $H_{S}$:
\begin{equation}
    \Delta R(H)=\Delta R(\infty)\frac{H^{2}}{H^{2}+H_{S}^{2}}
\end{equation}
A typical fit is seen in the 0.3 k$\Omega$ stage in fig. 2b.  The extracted saturated fields from these fits match those extracted from the Hall effect curves for all stages of evaporation.

\begin{figure}
{\epsfxsize=2.8 in \epsffile{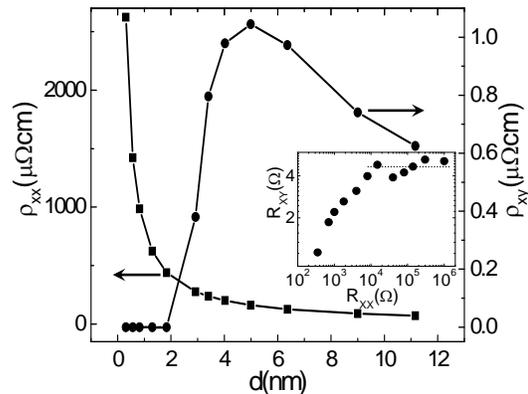}}
\vspace{-0.3cm}
\caption{ Longitudinal,
$\rho_{xx}$, and saturated transverse (Hall) resistance, $\rho_{xy}$
of a homogenous Ni sample versus the thickness of the films. The inset shows
the saturated transverse (Hall) resistance, $R_{xy}$ versus longitudinal
sheet resistance of a Ni \emph{granular} film.  \small}
\label{granular}
\vspace{-0.3cm}
\end{figure}

Hence both HE and MR measurements imply that for strong enough disorder, $R>R_{C}$, the
sample does not show signs of spontaneous magnetization and that
ferromagnetism emerges only for films characterized by
smaller resistances. One may wonder whether this effect can be due
to granularity in the film. While the samples are electrically
continuous for films with thicknesses of ~0.2 nm we can not rule out
the presence of some film granularity. For comparison we have
produced granular films of Ni with grain sizes of about 20 nm in
diameter and 2 nm in height \cite{aviad3}. These are achieved by quench condensing
Ni on a bare Si/SiO substrate without depositing a Sb or Ge
wetting layer \cite{aviad1, aviad2}. For the thinnest films of this
type the grains are superparamagnetic showing no hysteresis in the
magnetoresistance curve \cite{aviad1}. In the granular case, EHE is
observed even for the highest-resistance measured samples $R\sim1$
$M\Omega$. As material is added the resistance drops considerably,
however, the Hall effect resistance hardly changes until $R<10$
$k\Omega$, as can be seen in the inset of fig 3. Such behavior has
been also reported for granular Fe prepared by different methods
\cite{hebard} where it was understood that in the granular case the
EHE is due to scattering within the grains rather than being due to
hopping between grains. For $R<10$ $k\Omega$ $R_{xy}$ reduces
linearly with $R_{xx}$. This is the regime in which the grains are
expected to coalescing and the behavior approaches that of a
continuous film. In any case, a granular film  behaves very
differently than our uniform films.

We have also considered the possibility that appearance of the  magnetization is a function of the film thickness. A number of experiments on epitaxially grown Ni, Co and Fe thin films on Cu substrates \cite {d1,d2,d3,d4} have shown that the Curie temperature, $T_{C}$ drops sharply to zero for thicknesses approaching an atomic monolayer. These samples were ordered crystalline films and the substrates were metallic, hence, no effect of electron localization could be expected. The suppression of $T_{C}$ was interpreted as the formation of magnetic dead layers due to electronic hybridization with the states in the Cu substrate. This reduces the density of states and the Stoner criterion is defied. A similar conclusion was drawn by Bergman \cite{bergmann} who saw no signs for magnetization in the Hall effect of quench condensed Ni evaporated on an amorphous metallic $Pb_{75}Bi_{25}$ substrates for films thinner than $2.5$ monolayers. It seems that our results are different from all the above. We note that our films are grown on insulating substrates, hence electronic hybridization is not relevant. Consequently we are unable to detect magnetization in Ni films that are $1.8$ nm thick ($\sim7$ monolayers), which is much thicker than any of the previous experiments.
\begin{figure}
\vspace{-0cm}
{\epsfxsize=2.8 in \epsffile{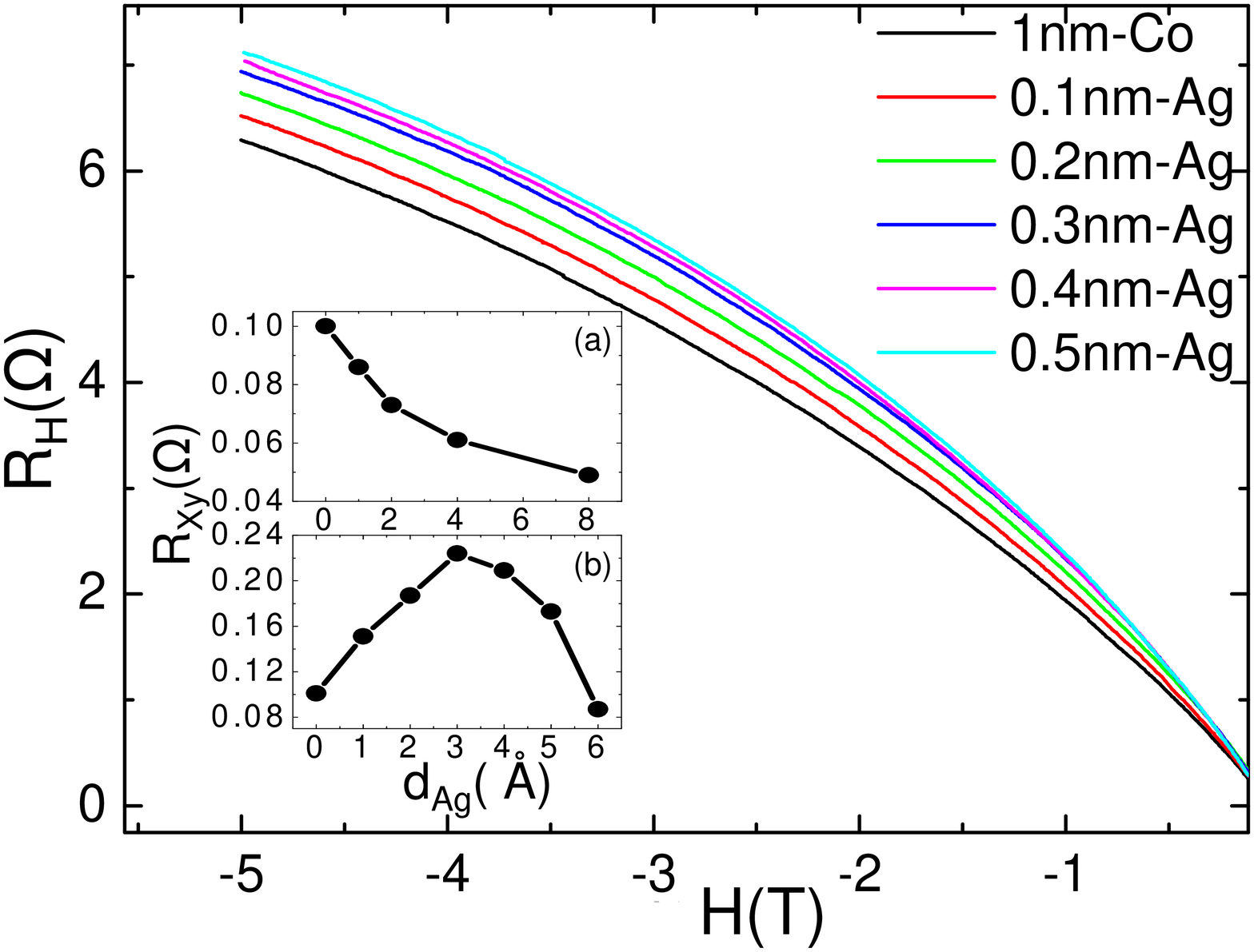}}
\vspace{-0.3cm}
\caption{ Hall effect measurements of a series of a 1.1 nm Co film coated by Ag overlayers. The inset shows the saturated transverse (Hall) resistance, $R_{xy}$ of a 2nm thick Ni film (a) and a 1nm thick Co sample (b) versus the thickness of the Ag overlayer.  \small}
\label{overlayer}
\vspace{-0.3cm}
\end{figure}

In order to test the notion that localization of the conduction electrons and not the thickness of the films is the cause for absence of magnetization we performed an experiment designed to vary the degree of localization without changing the ferromagnetic layer thickness. For this purpose we deposited thin layers of normal (non-magnetic) Ag on top of a thin ferromagnetic layer in the vicinity of $R_{C}$.  The normal metal is not expected to increase the amount of magnetic material but it can reduce electron localization and screen out electronic interactions. This is similar to an experiment performed on ultrathin layers of superconductors in which the critical temperature is suppressed by disorder \cite{inversePE,inversePE2}. Such films exhibit an "inverse proximity effect"; Addition of ultrathin normal metal layers acts to \emph{increase } $T_{C}$ rather than reduce it as expected from the proximity effect. In our ferromagnetic films we find a similar effect. For thick films, adding any amount of an Ag overlayer reduces the EHE signal. This is the case for Ni films for which $d_{C}\sim1.8$ nm. Indeed one expects that electronic hybridization with the Ag electrons should  suppress ferromagnetism. On the other hand, for 1 nm thick Co films, the first few Ag overlayers have an opposite effect. As can be seen in fig. 4, adding ultrathin normal metal layers plays the same role as adding ferromagnetic layers in that it \emph{increases} the EHE signal. Thick enough Ag overlayers result in a change of trend and suppression of the EHE. This is seen in the inset of fig 4 where we compare the outcome of adding ultrathin layers of Ag to 2 nm thick and 1 nm thick ferromagnetic films. The results of the 1nm Co film demonstrate that reducing the disorder without changing the ferromagnet thickness results in an increase in the magnetization despite the fact that there is a competing effect of electron hybridization.

In summary, our results show that no ferromagnetism can be detected by Hall effect or magnetoresistance measurements for films having resistances larger than $R_{C}$. We note that $R_{C}$, which is material dependent, seems to correlate with the material atomic magnetic moment $\mu$, as can be seen in table 1. Hence it is possible that the critical point at which magnetization appears in the film is related to the fraction of magnetization that is due to localized moments on the atoms. Clearly, more theoretical work is required to clarify the interplay between itinerant ferromagnetism and Anderson localization. Such research may prove to be very useful in shedding light on the old puzzle of the origin of magnetism in the transition metals.

We are grateful for fruitful discussions with D. Golosov, A.M. Goldman and R. Berkovits.  This research
was supported by the Israeli academy of science (grant number 249/05).

\end{document}